\documentclass[twocolumn]{aastex63} 
\usepackage{graphicx}
\usepackage{natbib}
\usepackage{placeins}
\usepackage{amsmath}
\usepackage{tgcursor}
\usepackage{booktabs}
\usepackage{enumitem}
\setcitestyle{notesep={ }}

\begin{document}

\title{First images of phosphorus molecules towards a proto-Solar analog}

\author[0000-0002-8716-0482]{Jennifer B. Bergner}
\altaffiliation{NASA Sagan Fellow}
\affiliation{University of Chicago Department of the Geophysical Sciences, Chicago, IL 60637, USA}

\author[0000-0003-0799-0927]{Andrew M. Burkhardt}
\affiliation{Wellesley College Department of Physics, Wellesley, MA  02481, USA}

\author[0000-0001-8798-1347]{Karin I. \"Oberg} 
\affiliation{Center for Astrophysics \textbar\ Harvard \& Smithsonian, 60 Garden St., Cambridge, MA 02138, USA}

\author[0000-0002-7231-7328]{Thomas S. Rice} 
\affiliation{Department of Astronomy, Columbia University, 550 West 120th Street, New York, NY 10027, USA}

\author[0000-0003-4179-6394]{Edwin A. Bergin} \affiliation{Department of Astronomy, University of Michigan, 1085 S. University Avenue, Ann Arbor, MI 48109, USA}

\begin{abstract}
\noindent The chemistry of phosphorus in star- and planet-forming regions is poorly understood, despite the central role of phosphorus in terrestrial biochemistry.  We present ALMA Band 3 and 4 observations of PO and PN towards the Class I protostar B1-a, representing the first spatially resolved observations of phosphorus carriers towards a Solar-type star forming region.  The phosphorus molecules emit from two distinct clumps, which coincide with regions where the protostellar outflow (traced by SiO) interacts with a filament of dense gas (traced by CCS).  Thus, the gas-phase phosphorus seems to originate from the shocking of dense interstellar clumps.  Based on the observed emission patterns, PO and PN appear to be daughter products of a solid phosphorus carrier with an intermediate volatility between ices and silicate grains.  Interstellar shocks may therefore play an important role in converting semi-refractory phosphorus to a more volatile form prior to incorporation into cometary ices.  Indeed, the (PO+PN)/CH$_3$OH ratio is similar in B1-a and comet 67P, implying a comparable reservoir of volatile phosphorus.  The PO/PN ratio ranges from $\sim$1--8 across B1-a.  The northern emission clump exhibits a lower PO/PN ratio and weaker $^{13}$CH$_3$OH emission than southern clump, indicating distinct shock physics and chemistry at the two positions.  Resolved observations of P carriers towards additional sources are needed to better understand what regulates such variations in the PO/PN ratio in protostellar environments.

\end{abstract}


\section{Introduction}
\label{sec:intro}
Phosphorus is an essential ingredient in the machinery of terrestrial life, playing a fundamental role in a diverse array of biochemical functions such as cell structure, genetics, and metabolism \citep{Macia2005}.  Because of this, understanding how phosphorus is delivered to nascent planetary surfaces is a key question in origins of life studies.  Characterizing the astrochemistry of phosphorus in low-mass protostellar regions is of particular importance in this regard, as these objects are the evolutionary progenitors of planetary systems like our own.  Indeed, the recent detection of phosphorus in the coma of comet 67P/Churyumov-Gerasimenko, hereafter comet 67P, \citep{Altwegg2016, Rivilla2020} enables direct comparisons between the volatile phosphorus inventory of the young Solar system with protosolar analogs.

To date, there have been only two detections of phosphorus molecules towards low-mass star-forming regions: PN and PO were detected towards the B1 shock position of the L1157 outflow \citep{Yamaguchi2011, Lefloch2016}, and towards the B1-a protostar \citep{Bergner2019}.  In both cases, the phosphorus molecule emission appears to be associated with outflow shocks.  Outflows are a common feature of protostellar environments, and so the scarcity of phosphorus detections in low-mass star forming regions indicates that the presence of an outflow is on its own insufficient to produce abundant gas-phase PO and PN.  Currently, it is unclear what additional element is needed.  A comparison of the PO and PN kinematics in B1-a and L1157 do not reveal any obvious trends: the phosphorus lines are a factor of $\sim$4 broader in L1157 compared to B1-a, with wings extending to velocities $\sim$5$\times$ higher, suggestive of emission from a more energetic environment \citep{Lefloch2016, Bergner2019}.

The detection of phosphorus molecules in massive star forming regions similarly appears to be correlated with the presence of shocked material \citep{Mininni2018, Rivilla2018, Fontani2019, Rivilla2020, Bernal2021}.  This consistent association across dense interstellar environments indicates that phosphorus is normally stored in the solid state, and some form of grain sputtering is responsible for releasing it into the gas within shocked regions.  Indeed, the lack of phosphorus molecule emission towards hot core environments indicates that ice sublimation is not responsible for populating the gas with phosphorus \citep{Rivilla2020, Bernal2021}.  \citet{Rivilla2020} also proposed that, along with shocks, photochemistry is an important ingredient in producing gas-phase P molecules.  Curiously, in all dense interstellar regions where PO has been detected, the measured PO/PN ratio falls within the range of $\sim$1--8 \citep{Lefloch2016, Rivilla2016, Rivilla2018, Bergner2019, Rivilla2020, Bernal2021, Haasler2021}.  It remains unknown why the PO/PN ratio is so consistent across different star-forming environments, though it may be related to a similar chemistry and physics at play within the various shocked environments. 

A major limitation in characterizing the phosphorus chemistry along the star formation sequence is that most previous detections of PO and PN were made with single-dish telescopes.  Without spatial information, it is not possible to (i) derive precise column densities or abundances due to unknown beam dilution factors, (ii) measure variations in column densities across the source, or (iii) compare emission morphologies of different line tracers.  The recent mapping of PO and PN towards the massive star-forming region AFGL 5142 with ALMA was a major step forward in this regard \citep{Rivilla2020}.  Here, we present the first maps of phosphorus molecules towards a low-mass star forming region, the Class I protostar B1-a located in the Perseus molecular cloud \citep[distance$\sim$300 pc; ][]{Ortiz2018}.  Section \ref{sec:obs} describes our ALMA observations.  In Section \ref{sec:obs_results} we present the PO and PN emission maps and compare their morphologies with other molecular tracers.  In Section \ref{sec:columns} we derive maps of the molecular column densities and column density ratios.  In Section \ref{sec:discussion} we discuss the implications of our findings with respect to the chemistry of phosphorus in low-mass star forming regions.  Section \ref{sec:concl} summarizes our main conclusions.

\section{Observations}
\label{sec:obs}
Observations of B1-a were taken during ALMA Cycle 7 as part of the project 2019.1.00708.S (PI: J. Bergner).  The spectral setup was designed to cover transitions of PO and PN in both Band 3 and Band 4.  Each setup was observed with both a low- and high-resolution antenna configuration, resulting in angular resolutions around 1$\arcsec$, and a maximum recoverable scale around 25$\arcsec$ for Band 3 and 38$\arcsec$ for Band 4.  The phase center of the observations was the B1-a continuum source (J2000 R.A.=03:33:16.67, Decl.=+31:07:55.1).  Along with the phosphorus molecules, our observations covered transitions of the molecular tracers SO$_2$, $^{13}$CH$_3$OH, and CCS.  Additional details on the correlator configurations can be found in Appendix \ref{sec:app_obsdeets}.  The line data for the transitions used in this work can be found in Table \ref{tab:line_dat}. 
\begin{deluxetable}{lccccc}
	\tabletypesize{\footnotesize}
	\tablecaption{Spectral line data \label{tab:line_dat}}
	\tablecolumns{6} 
	\tablewidth{\textwidth} 
	\tablehead{
        \colhead{Molecule}       & 
        \colhead{Frequency}       & 
        \colhead{Transition} &
		\colhead{log($A_{ul}$)} & 
		\colhead{$E_u$}       &
		\colhead{$g_u$}          \\
		\colhead{} & 
		\colhead{(GHz)} &
        \colhead{} &    
		\colhead{ (s$^{-1}$)} &
		\colhead{(K)} &
		\colhead{} }
\startdata
\hline 
PN & 93.978206 & N=2--1, J=2--2 & -5.14 & 6.8 & 5 \\
 & 93.978473 & N=2--1, J=1--0 & -4.79 & 6.8 & 3 \\
 & 93.979769 & N=2--1, J=2--1 & -4.66 & 6.8 & 5 \\
 & 93.979890 & N=2--1, J=3--2 & -4.54 & 6.8 & 7 \\
 & 93.982319 & N=2--1, J=1--1 & -4.92 & 6.8 & 3 \\
 & 140.966008 & N=3--2, J=3--3 & -4.93 & 13.5 & 7 \\
 & 140.967425 & N=3--2, J=2--1 & -4.05 & 13.5 & 5 \\
 & 140.967692 & N=3--2, J=3--2 & -4.03 & 13.5 & 7 \\
 & 140.967764 & N=3--2, J=4--3 & -3.98 & 13.5 & 9 \\
 & 140.969975 & N=3--2, J=2--2 & -4.79 & 13.5 & 5 \\
\hline
PO & 108.998445 & $a$, F= 3--2, l=e & -4.67  & 8.4 & 7 \\
& 109.045396 &    $a$, F= 2--1, l=e & -4.72  & 8.4 & 5 \\
& 109.206200 &    $a$, F= 3--2, l=f & -4.67  & 8.4 & 7 \\
& 109.271376 &    $a$, F= 2--2, l=e & -5.67  & 8.4 & 5 \\
& 109.281189 &    $a$, F= 2--1, l=f & -4.72  & 8.4 & 5 \\
& 152.656979 &    $b$, F= 4--3, l=e & -4.20  & 15.7 & 9 \\
& 152.680282 &    $b$, F= 3--2, l=e & -4.22  & 15.7 & 7 \\
& 152.855454 &    $b$, F= 4--3, l=f & -4.20  & 15.8 & 9 \\
& 152.888128 &    $b$, F= 3--2, l=f & -4.22  & 15.7 & 7 \\
\hline
$^{13}$CH$_3$OH & 141.595477 & 3$_{0,3}$--2$_{0,2}$ & -4.94  & 26.7 & 7 \\
& 141.597059 & 3$_{-1,3}$--2$_{-1,2}$ & -4.99  & 19.2 & 7 \\
& 141.602528 & 3$_{0,3}$--2$_{0,2}$ ++ & -4.94  & 13.6 & 7 \\
& 141.629262 & 3$_{1,2}$--2$_{1,1}$ & -4.98  & 34.6 & 7 \\
\hline 
SiO & 86.84696  & 2--1 & -4.53  & 6.2 & 5 \\
    & 260.51802 & 6--5 & -3.04 & 43.8 & 13 \\
CCS & 93.870107 & N=7--6, J=8--7 & -4.43  & 19.9 & 17 \\
SO$_2$ & 140.30617 & 6$_{2,4}$--6$_{1,5}$ & -4.60  & 29.2 & 13 \\
\enddata
\tablenotetext{}{$a$: J=$\frac{5}{2}-\frac{3}{2}$, $\Omega=\frac{1}{2}$, $b$: J=$\frac{7}{2}-\frac{5}{2}$, $\Omega=\frac{1}{2}$.  Line parameters are taken from the CDMS catalog \citep{Muller2001,Muller2005} based on data from \citet{Cazzoli2006} (PN), \citet{Bailleux2002} (PO), \citet{Xu1997} ($^{13}$CH$_3$OH), \citet{Muller2013} (SiO), \citet{Saito1987} (CCS), and \citet{Muller2005b} (SO$_2$).}
\end{deluxetable}

\begin{figure*}
\centering
    \includegraphics[width=\linewidth]{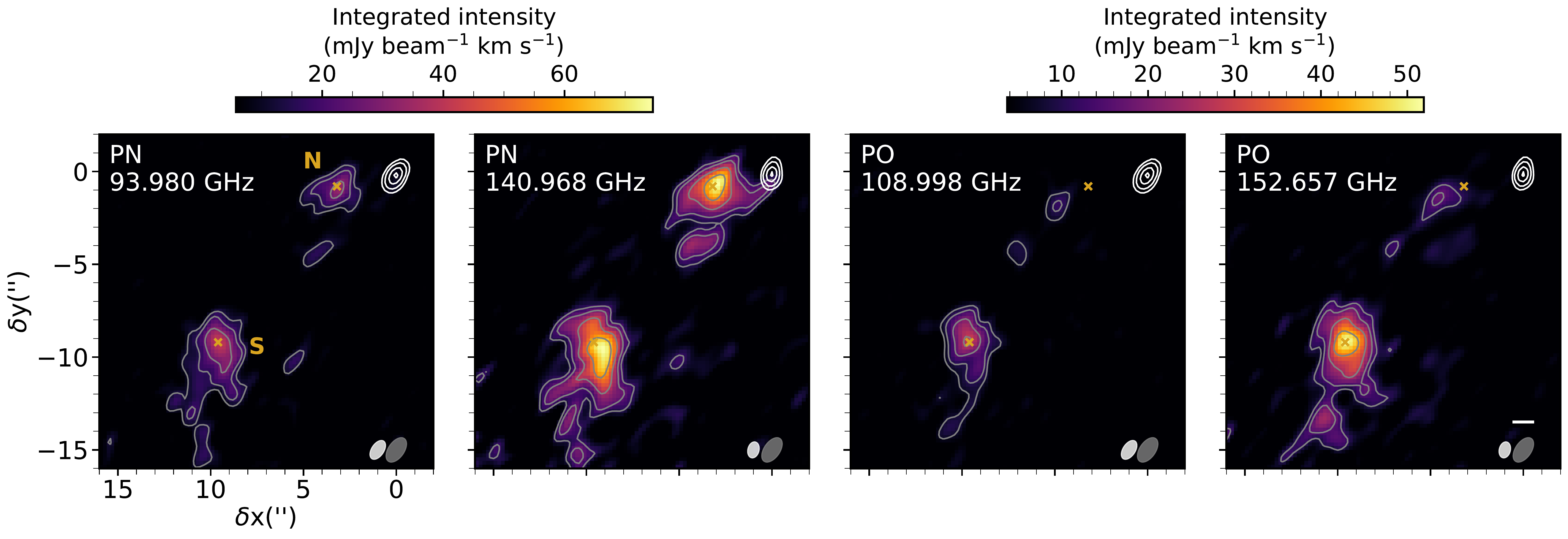}
    \caption{Moment zero maps of Band 3 and Band 4 PN (left) and PO (right) emission towards B1-a.  Colorbars are shared for both transitions of each molecule.  Grey contours represent [4,6,10,15]$\sigma$ levels for the line data.  White contours show the [10,20,40,60]$\sigma$ continuum contours, where $\sigma$ is 0.03 and 0.06 mJy beam$^{-1}$ for the continuum B3 and B4 images, respectively.  The restoring beams for the line and continuum images are shown in grey and white, respectively, in the bottom right of each panel.  The north and south clump are labeled `N' and `S', and gold $\times$'s mark the positions used to extract spectra (Figure \ref{fig:spec_compare}).  The scale bar in the right panel represents a linear distance of 300 au.}
    \label{fig:obs_summary}
\end{figure*}

The dataset was calibrated using the ALMA pipeline with CASA \citep{McMullin2007} version 5.6.1-8.  The same CASA version was used for all subsequent analysis.  Individual execution blocks were imaged separately to check for consistency in the pointing and calibration.  No self-calibration solutions were applied as it was not found to meaningfully improve the SNR.  Continuum subtraction was performed on the line data using the \texttt{uvcontsub} task.  The relative visibility weights were checked prior to combining the long- and short-baseline data for imaging.  

Imaging was performed using the \texttt{tclean} task in CASA.  We used the multi-scale clean algorithm with scales of [0, 1.6$\arcsec$, 4$\arcsec$].  Briggs weighting was used with a robust value of 0.0 for all lines except $^{13}$CH$_3$OH, for which we adopted a value of 1.0.  Images were generated with a velocity spacing of 0.4 km s$^{-1}$ for Band 3 and 0.3 km s$^{-1}$ for Band 4.  Clean masks were drawn by hand to include the observed emission.  We cleaned to a threshold of 4$\sigma$, where $\sigma$ is the average channel rms across 10 line-free channels in the dirty image.  We used the CASA task \texttt{imsmooth} on the Band 4 PO and PN images in order to achieve the same beam dimensions as the corresponding Band 3 images, since beam-matched images are needed for deriving PO and PN column densities. The resulting image properties for the spectral windows used in this work are listed in Table \ref{tab:image_dat}.

We also make use of observations covering the SiO 2--1 and 6--5 transitions taken with the NOEMA interferometer.  Details on these observations can be found in \citet{Bergner2019}.

\section{Observational results}
\label{sec:obs_results}
\subsection{PO and PN emission}

Our observations cover numerous hyperfine transitions of PO and PN in ALMA Bands 3 and 4, listed in Table \ref{tab:line_dat}.  Figure \ref{fig:obs_summary} shows the moment-zero maps of the strongest PN and PO transitions in both Band 3 and Band 4.  The millimeter continuum traces a compact dusty structure in the protostellar core.  The PN and PO emission is offset from the continuum peak, with a prominent clump slightly east of the continuum peak, and another clump about 15$\arcsec$ southeast of the continuum peak.  In subsequent analysis, we refer to these as the north (N) clump and south (S) clump, respectively.  PN emits brightly from both regions, while PO emits only weakly from the N clump.  As traced by the Band 4 PN transition, both emission clumps span $\sim$5$\arcsec$ (1500 au) in diameter, and are well-resolved by our observations.

We report source-integrated line fluxes by summing the emission from all pixels with a SNR$>$3 in the moment zero map.  The resulting fluxes, along with the corresponding emitting areas, are listed in Table \ref{tab:image_dat}.  Integrated flux uncertainties are found by bootstrapping: we apply the same spatial mask to 100 moment zero maps made from randomly selected line-free channels, and solve for the standard deviation across the resulting integrated fluxes.

\begin{figure*}
\centering
    \includegraphics[width=\linewidth]{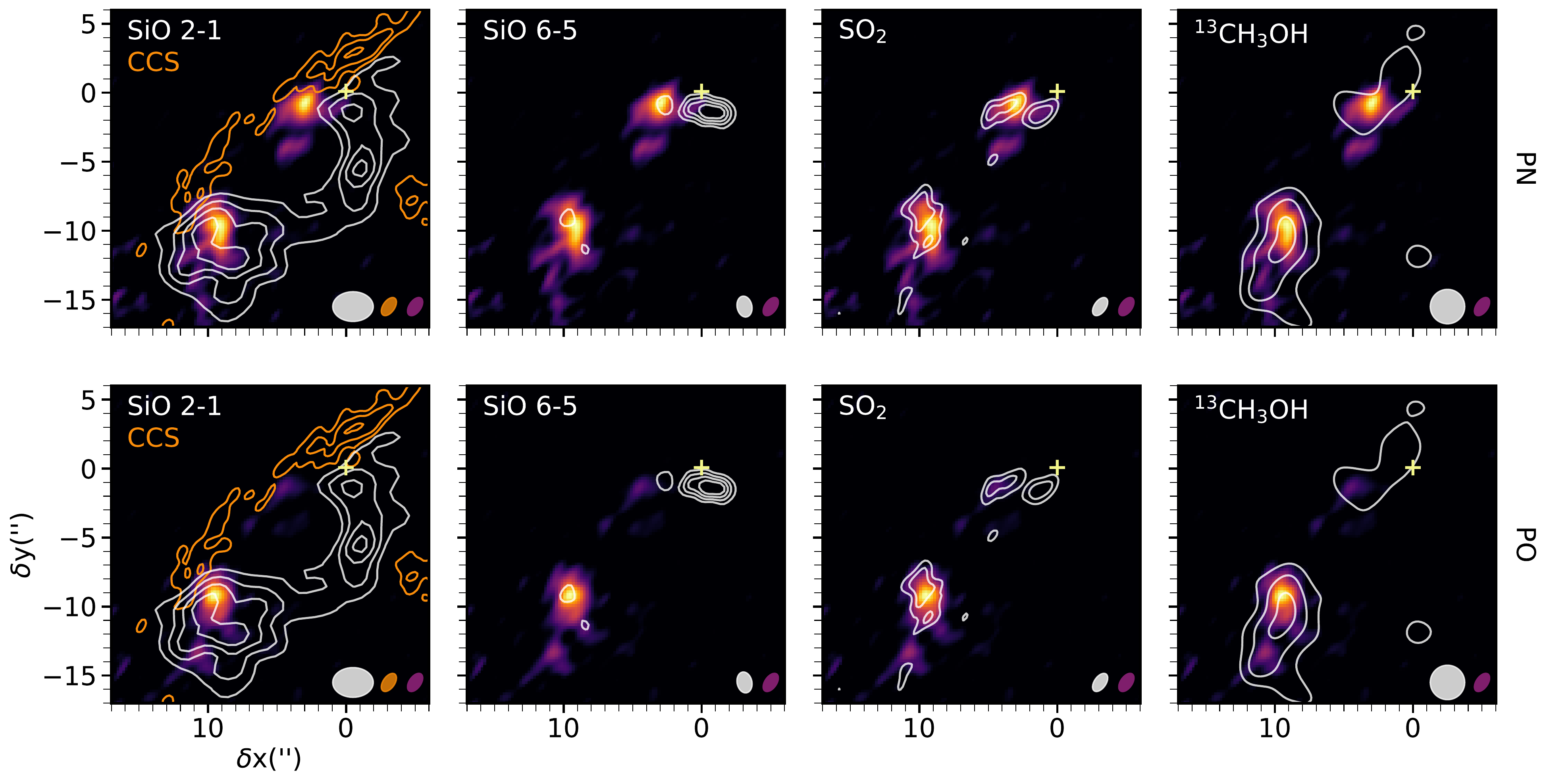}
    \caption{Emission morphologies of Band 4 transitions of PN and PO (colormaps), compared with SiO, CCS, SO$_2$, and $^{13}$CH$_3$OH.  Contour levels correspond to [4,7,10,13]$\sigma$ for the line data, where $\sigma$ is 54, 56, 1.7, 3.6, and 3.2 mJy beam$^{-1}$ km s$^{-1}$ for SiO 2--1, SiO 6--5, CCS, SO$_2$, and $^{13}$CH$_3$OH, respectively.  The continuum peak position is shown with a yellow `+'.  Restoring beams for PN and PO (purple) and the additional lines (grey/orange) are shown in the bottom right of each panel.}
    \label{fig:emission_compare}
\end{figure*}

\subsection{B1-a outflow morphology}
\label{subsec:outflow_tracers}
We now present the emission maps of other molecular line tracers in order to explore the origin of phosphorus molecule emission within the B1-a protostar.  Figure \ref{fig:emission_compare} shows the moment zero maps of SiO, CCS, SO$_2$, and $^{13}$CH$_3$OH transitions compared to the PN (top) and PO (bottom) emission.  

SiO is a well established tracer of molecular outflows around forming stars \citep{Martin-Pintado1992}.  While Si is usually heavily depleted from the gas in the dense ISM, its gas-phase abundance is enhanced by orders of magnitude within outflows.  This may be due to shock-induced dust destruction \citep[e.g.][]{Schilke1997, Guillet2009, Guillet2011}, or alternatively the launching of the jet from within the dust sublimation zone near the protostar \citep{Glassgold1991}.  In B1-a, the SiO 2--1 transition (E$_\mathrm{up}$=6 K) traces an apparent outflow south and then east from the dust continuum (Figure \ref{fig:emission_compare}).  The outflow as traced by SiO appears to be monopolar, with no SiO emission detected to the north of the continuum center, which could reflect asymmetric outflow ejections.  The position of the outflow cavity is not clear from these observations.  Additional line tracers are needed to better understand the nature of the B1-a outflow, since there is precedent for monopolar SiO emission accompanying a bipolar CO outflow \citep{Codella2014}.  While the S clump of PO and PN emission overlaps with the SiO outflow, phosphorus molecules do not emit co-spatially with the rest of the SiO outflow.  Indeed, the phosphorus N clump is adjacent to but not overlapping with the SiO outflow.  Thus, phosphorus molecules do not appear to be localized to the outflow as traced by SiO.

\begin{figure}
\centering
    \includegraphics[width=\linewidth]{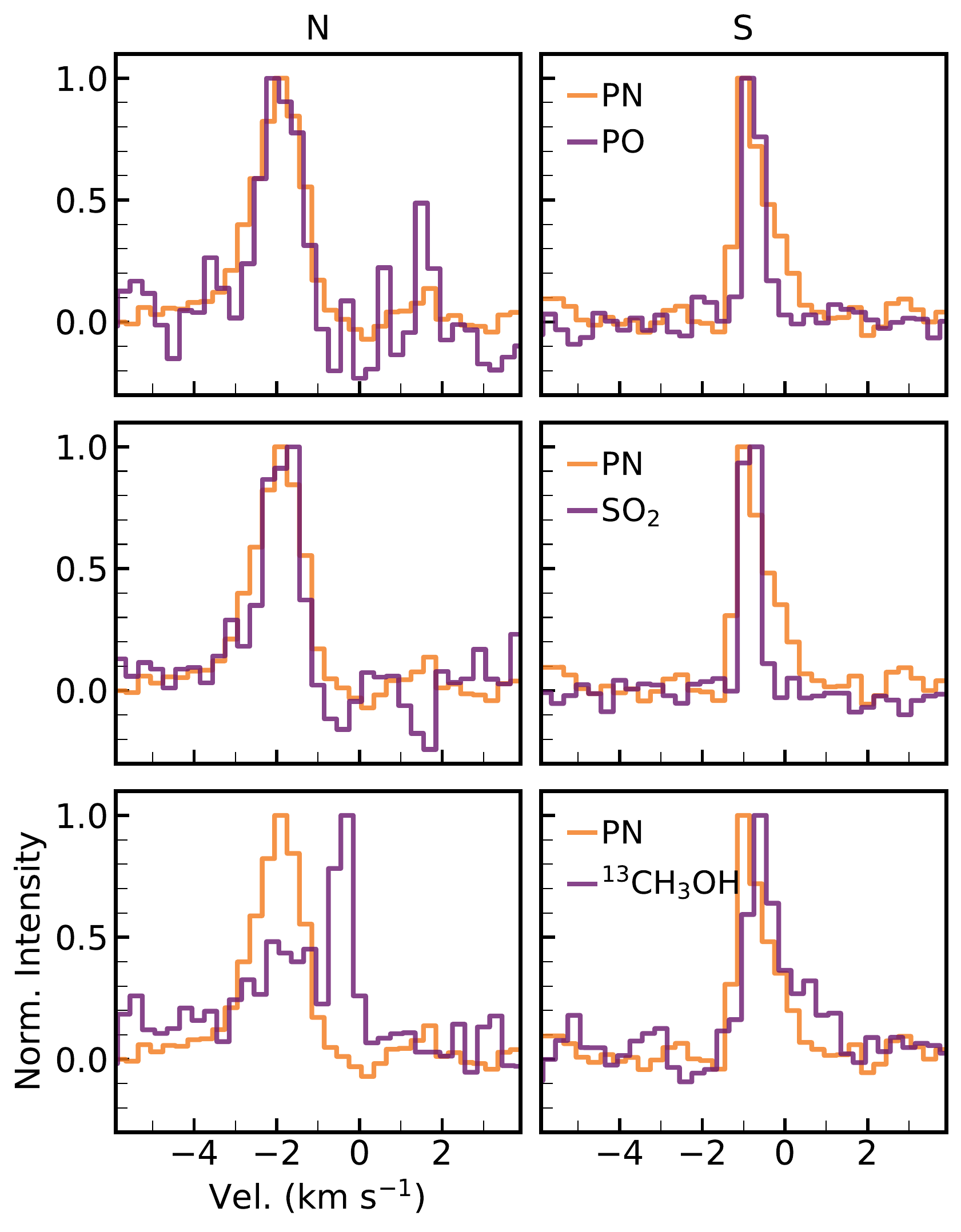}
    \caption{Normalized line profiles of PN, PO, SO$_2$, and $^{13}$CH$_3$OH.  For reference, PN is shown in each panel.  Spectra were extracted from a 5$\times$5 pixel square centered on the N and S positions marked in Figure \ref{fig:emission_compare}.}
    \label{fig:spec_compare}
\end{figure}
Interestingly, the two phosphorus emission peaks occur at the two locations where the SiO outflow intersects with a filament-like structure traced by CCS J$_\mathrm{N}$=8$_7$--7$_6$.  CCS is considered an indicator of `chemically young' material \citep{Suzuki1992}, and has been shown to trace the edges of filaments or clumps of dense gas within star-forming cores \citep[e.g.][]{Langer1995, Velusamy1995, Kuiper1996, Peng1998, Lai2003, Dobashi2019}.  Taken together with the elongated CCS morphology towards B1-a and the high critical density of the 93 GHz transition ($\sim$7$\times$10$^{6}$ cm$^{-3}$), we suspect it traces a small-scale, dense, and chemically/dynamically young filament.  It appears that phosphorus is released into the gas where the SiO outflow encounters this dense clumpy material, and not within the SiO outflow itself.  Still, given the unusual morphology of the SiO outflow and the unknown position of the outflow cavity, it remains possible that the PO and PN emission clumps are also associated with the cavity wall as was seen in AFGL 5142 \citep{Rivilla2020}.

The SiO 6--5 line (E$_\mathrm{up}$=44 K) should trace warmer gas than the 2--1 line.  While SiO 6--5 mainly emits just south of the continuum center in a region without PO or PN emission, it also exhibits small emission peaks overlapping with the N and S clumps of phosphorus emission.  Interestingly, it appears to peak co-spatially with PN in the N clump and with PO in the S clump.  Thus, neither PN nor PO show a clear association with the higher-excitation SiO.

SO$_2$ and CH$_3$OH are both considered tracers of low-velocity shocks, and become abundant in the gas phase due to ice sputtering or sublimation at velocities or temperatures lower than those required to disrupt the refractory grain \citep[e.g.][]{Blake1995, Bachiller1997}.  Note that CH$_3$OH should desorb directly from grains, whereas SO$_2$ may instead be a product of gas-phase chemistry upon the release of another (unknown) S carrier \citep{Bachiller1997, Jimenez2005}.  Figure \ref{fig:emission_compare} shows the emission morphology of the SO$_2$ 6$_{2,4}$--6$_{1,5}$ and $^{13}$CH$_3$OH 3$_{0,3}$--2$_{0,2}$ transitions.  We focus on the $^{13}$C isotopologue of CH$_3$OH because the rarer isotopologue is a better tracer of the outflow-associated material whereas the main isotopologue also traces envelope emission.  SO$_2$ and $^{13}$CH$_3$OH are both broadly similar in their emission morphologies to the phosphorus molecules, with emission peaks at the N and S positions of the phosphorus clumps.  This similarity suggests that, like SO$_2$ and CH$_3$OH, the phosphorus molecules are released into the gas due to weak shocks.  We also highlight that the distribution of SO$_2$ more closely resembles that of PN than PO, with comparable emission intensities in the N and S clumps.  $^{13}$CH$_3$OH is more similar in morphology to PO, with much weaker emission at the N clump compared to the S clump.  

\subsection{Line kinematics}
\label{subsec:lineprofiles}

\begin{figure*}
\centering
    \includegraphics[width=0.9\linewidth]{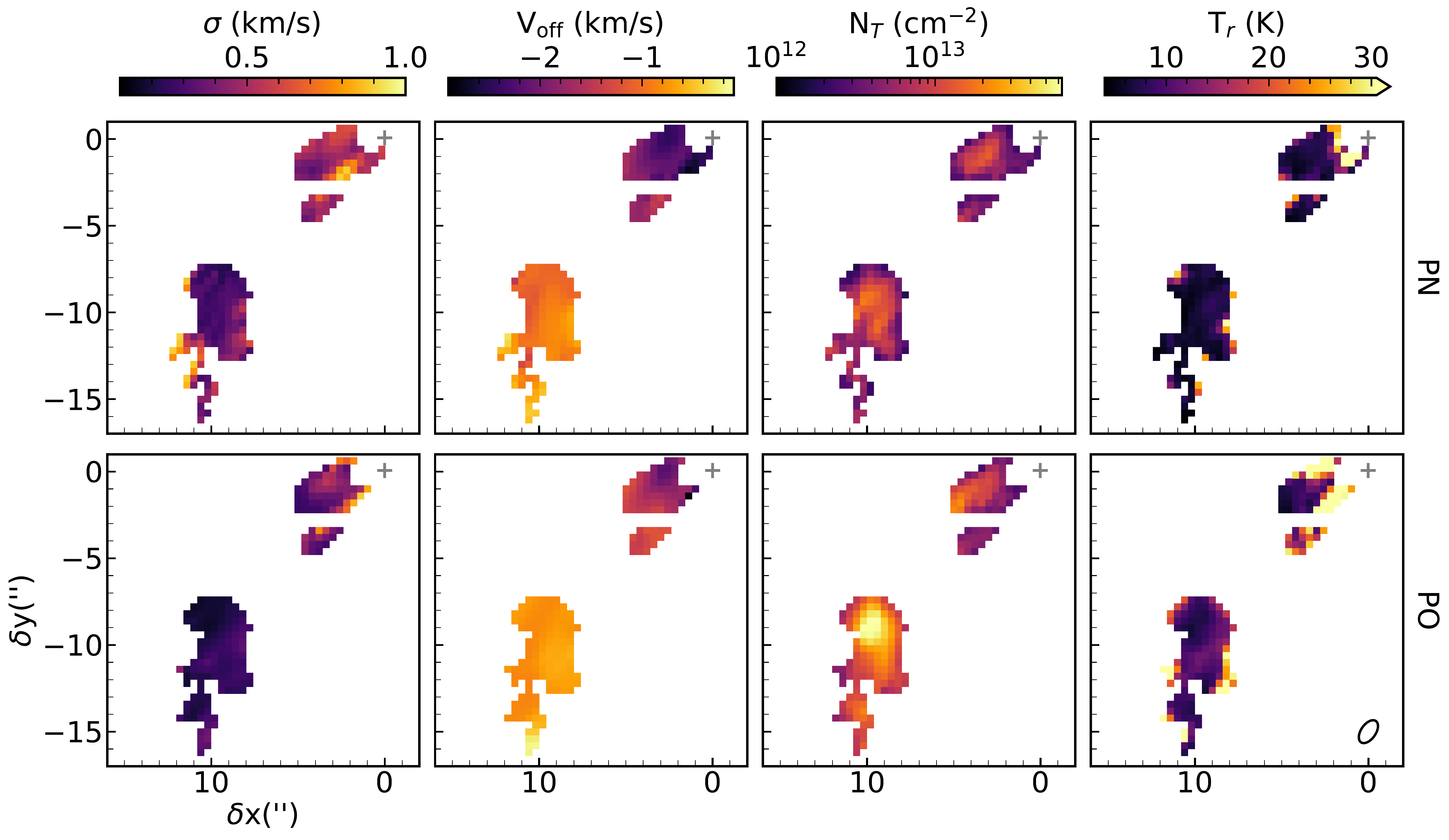}
    \caption{Maps of the best-fit parameters (i.e.~50th percentile posteriors) from spectral line fitting of PN (top) and PO (bottom).  Only pixels with $>$3$\sigma$ constraints on the column density are shown.  The restoring beam (common to PN and PO) is shown in the bottom right panel.}
    \label{fig:fit_parameters}
\end{figure*}

To further explore the relationship between the co-spatial molecules SO$_2$, $^{13}$CH$_3$OH, PO, and PN, Figure \ref{fig:spec_compare} shows the spectra extracted from each emission clump, at the positions marked with gold $\times$'s in Figure \ref{fig:obs_summary}.  At both positions, the line profiles of PN, PO, and SO$_2$ are quite similar, with comparable velocity offsets and line widths.  Of these three, PN exhibits slightly broader lines (see also Section \ref{subsec:po_pn_fits}).  

$^{13}$CH$_3$OH exhibits one velocity component that overlaps with the phosphorus molecules, and an additional component red-shifted relative to the phosphorus lines.  Towards the N clump this feature is a separate emission peak centered around the source rest velocity, and towards the S clump it appears as a broad shoulder on the main emission line.  Thus, while $^{13}$CH$_3$OH shows some similarities with the phosphorus lines, it appears to emit from a broader range of physical environments.  The excess emission at velocities close to the source rest velocity may reflect an emission component from the envelope in addition to the phosphorus clumps.  

Comparing the two positions, the lines at the N clump are characterized by a slightly higher velocity offset relative to the source velocity, though the velocity is still only around -2 km s$^{-1}$.  Together with the narrow line widths ($<$1 km s$^{-1}$ within individual pixels across the emission map; Figure \ref{fig:fit_parameters}), this indicates that the emitting gas is fairly quiescent in both the N and S clumps.

\section{Column densities}
\label{sec:columns}
For PO, PN, and $^{13}$CH$_3$OH, the coverage of multiple spectral features provides leverage for constraining the rotational temperatures and column densities.  We describe our line fitting routine in Appendix \ref{sec:app_linefits}.  Note that we use an LTE spectral model, since we recover similar column densities compared to RADEX modeling (Appendix \ref{sec:app_nonlte}).  However, a non-LTE solution remains possible, and coverage of higher-J transitions is needed to firmly discriminate between LTE and non-LTE emitting conditions.  For the results presented in Sections \ref{subsec:po_pn_fits}--\ref{subsec:vs_lw}, we use the PO and PN images with a 1.5"$\times$0.85" beam (the same as in Figures \ref{fig:obs_summary} and \ref{fig:emission_compare}), regridded by a factor of 2 in RA and Dec. in order to reduce the number of pixels being fitted.  For obtaining (PO+PN)/CH$_3$OH ratios in Section \ref{subsec:p_ch3oh_ratio}, we fitted PO, PN, and $^{13}$CH$_3$OH images all tapered to a circular 2.5" beam in order to obtain sufficient SNR for fitting the $^{13}$CH$_3$OH lines.

\subsection{PO and PN fits}
\label{subsec:po_pn_fits}
Maps of the PN and PO line widths, velocity offsets, column densities, and rotational temperatures derived from spectral line fitting are shown in Figure \ref{fig:fit_parameters}.  Note that we show only pixels with a well-constrained column density fit, as defined in Appendix \ref{sec:app_linefits}.  Analogous maps showing the fit parameter uncertainties are shown in Figure \ref{fig:fit_param_errors}.  PO generally exhibits a narrower line-width compared to PN, particularly in the S clump.  As seen in Figure \ref{fig:spec_compare}, within a given pixel the velocity offsets of the PO and PN lines are quite similar.  For both  molecules, higher velocity offsets are seen within the N clump compared to the S clump.  Within each emission clump, the PO and PN column densities tend to peak in the interior and fall off towards the edges.  The rotational temperatures are lowest around the column density peaks, and appear to increase towards the edges of each clump.  We note that the rotational temperatures are not very well constrained due to the fairly narrow range in upper-state energies of the lines covered by our observations ($\sim$7-15 K; Table \ref{tab:line_dat}).  Since a trend of increasing temperatures towards the clump edges is independently recovered for both PN and PO, this suggests that the pattern is indeed real. Still, observations of higher-excitation PO and PN lines are needed to confirm the rotational temperatures shown here.

\begin{figure}
\centering
    \includegraphics[width=0.9\linewidth]{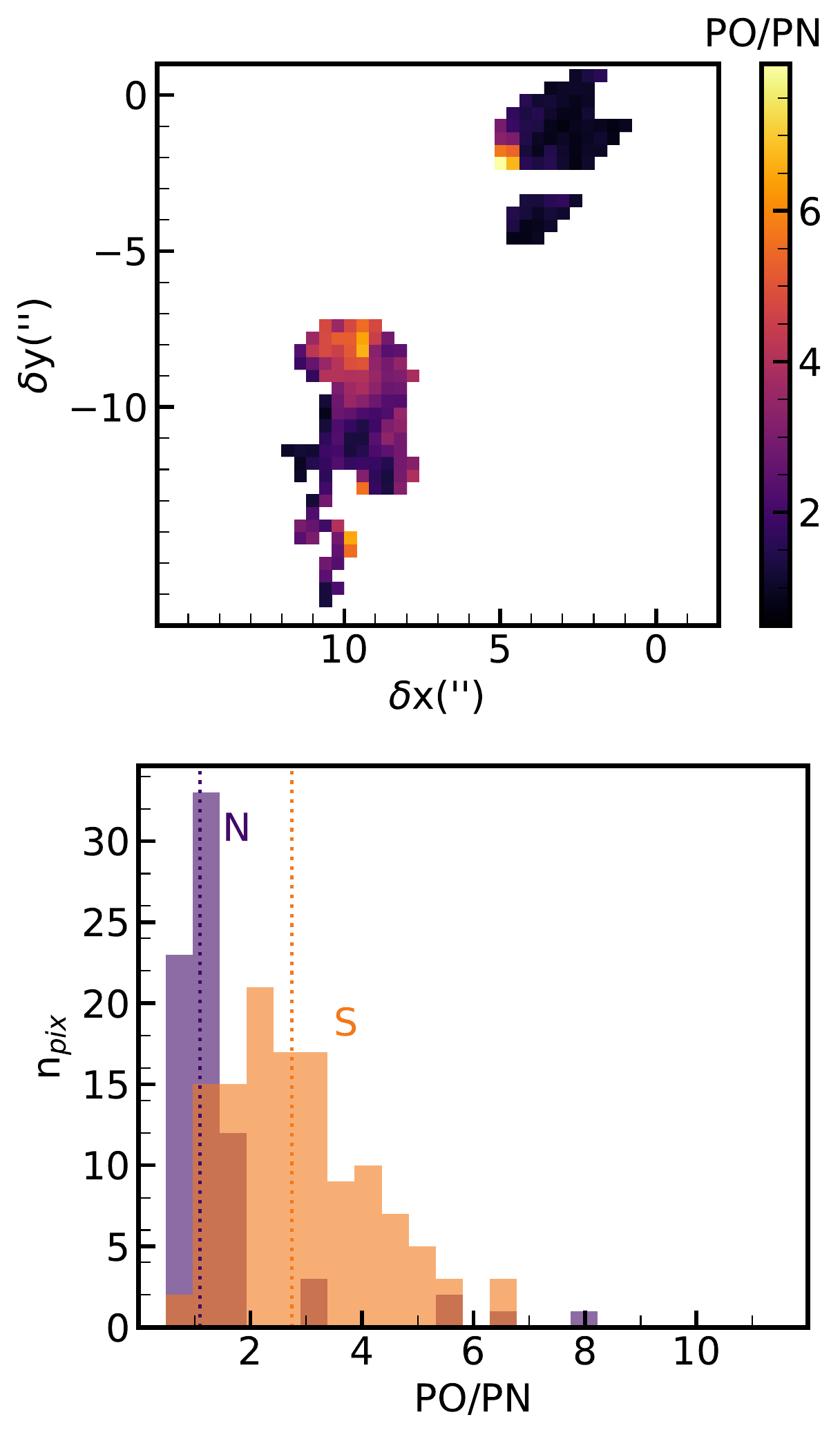}
    \caption{Top: map of the PO/PN ratio in B1-a.  Bottom: histogram of PO/PN ratios within the N and S clumps.  Median values for the N and S clumps (1.1 and 2.7, respectively) are shown with dotted lines.  For both panels, only pixels with well-constrained column densities of PN and PO are shown.}
    \label{fig:PO_PN_ratio}
\end{figure}

\subsection{PO/PN ratios}
\label{subsec:po_pn_ratio}
With spatially resolved column density maps, we can explore how the PO/PN ratio varies within the B1-a protostellar environment.  Figure \ref{fig:PO_PN_ratio} (top) shows the map of PO/PN ratios, where only pixels with well-constrained column densities for both PO and PN are included.  We note that the typical per-pixel uncertainty in PO/PN is around 30\%; individual pixel uncertainties can be seen in Figures \ref{fig:lw_columns}--\ref{fig:popn_vs_r}.  The distribution of PO/PN ratios within each clump is shown in Figure \ref{fig:PO_PN_ratio} (bottom).  Across the source, the PO/PN ratio ranges from $\sim$1--8, with a median of 1.9. The N clump generally exhibits low PO/PN ratios, with a median of 1.1, and most pixels concentrated in the range 1--2.  The S clump typically has higher PO/PN ratios with a median of 2.7, and shows more variation in PO/PN ratios compared to the N clump.  A few spots within each clump exhibit PO/PN ratios $>$6, though such high values are uncommon.

\subsection{Trends with line width}
\label{subsec:vs_lw}

\begin{figure*}
\centering
    \includegraphics[width=\linewidth]{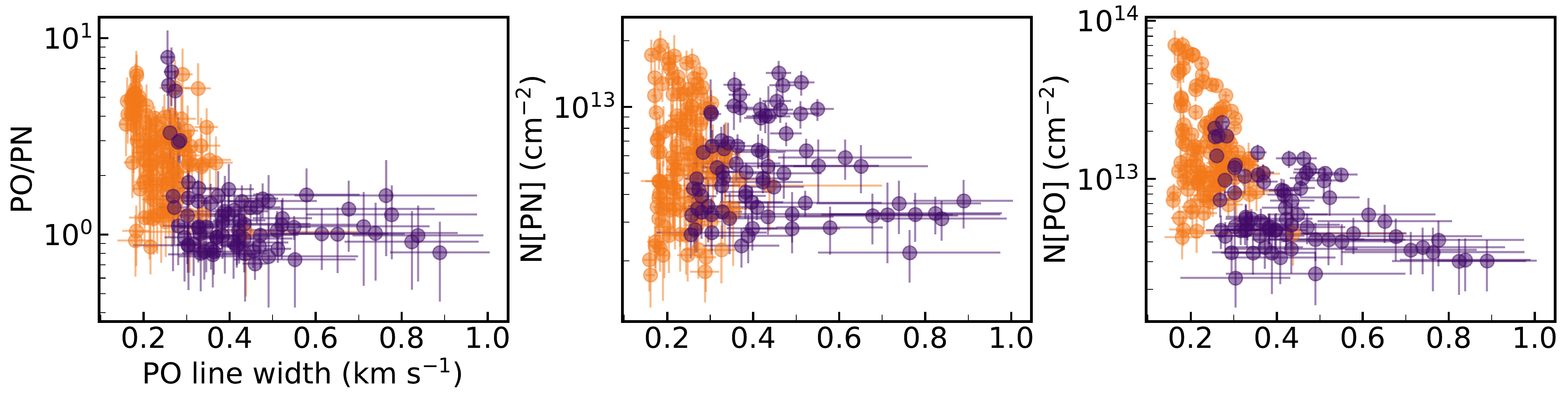}
    \caption{PO/PN, PN, and PO column densities within individual pixels as a function of PO line width.  Orange and purple colors correspond to the S and N clumps, respectively.}
    \label{fig:lw_columns}
\end{figure*}

To provide insight into the origins of PO/PN variations within B1-a, Figure \ref{fig:lw_columns} shows the PO/PN ratio as a function of the PO line width within a given pixel.  A similar trend is seen when the PN line width is used instead.  While we do not observe a monotonic trend between PO/PN ratio and line width, there is a clear clustering in which the PO/PN ratio is roughly constant around 1--2 for pixels with a PO line width $\gtrsim$0.3 km s$^{-1}$, and increases to $\sim$2--8 in regions with narrower line widths.  We can see from the middle and right panels of Figure \ref{fig:lw_columns} that this trend is predominantly driven by an increase in the PO column density for narrower line widths, whereas the PN column densities do not show any systematic variation with line width.

Spectral lines may be broadened as a result of higher temperatures or gas motions such as turbulence.  We identify a similar trend in which the PO/PN ratio is highest in regions with the lowest rotational temperatures, but focus our comparison on the line width since it is better-constrained (Section \ref{subsec:po_pn_fits}).  That PO is most abundant in regions with narrower linewidths suggests that its production may be favored in cooler and/or more quiescent gas compared to PN.  We discuss this further in Section \ref{sec:discussion}.  Spatially resolved observations of PO and PN in additional sources are needed to explore whether high PO/PN ratios in more quiescent regions is a general feature of the interstellar phosphorus chemistry, or unique to B1-a.  

\subsection{(PO+PN)/CH$_3$OH ratios}
\label{subsec:p_ch3oh_ratio}
In Section \ref{subsec:outflow_tracers} we showed that PO and PN emit co-spatially with the ice sublimation tracers $^{13}$CH$_3$OH and SO$_2$, and thus likely trace material sputtered into the gas by low-velocity shocks.  Here, we compare the gas-phase (PO+PN)/CH$_3$OH ratios measured in B1-a to the same ratio inferred for the Solar System comet 67P, in order to explore how the volatile P inventory of the early Solar System compares to a proto-Solar analog.  

We note several caveats to this comparison: first, it assumes that the parent phosphorus carriers are desorbed from grains with similar efficiencies as CH$_3$OH, i.e.~that the gas-phase (PO+PN)/CH$_3$OH ratio in B1-a is a proxy for the ice-phase ratio.  Second, it assumes that PO and PN are the dominant volatile P carriers.  In comet 67P, PO is at least 10$\times$ and 3$\times$ more abundant than PN and PH$_3$, respectively \citep{Rivilla2020}.  However, a contribution from elemental P could not be excluded based on the ROSINA data \citep{Rubin2019b, Rivilla2020}, and so it is possible that the true volatile P/CH$_3$OH abundance in comet 67P is higher than the value adopted here.  In the case of B1-a, shock chemistry models predict that PH$_3$ and P can in some circumstances have comparable abundances to PN and PO, depending on the physical conditions \citep{Jimenez2018}.  To date there are no constraining measurements of these carriers in a star-forming region.  Thus, the P/CH$_3$OH ratios we report for B1-a may also be lower limits.

\begin{figure}
\centering
    \includegraphics[width=\linewidth]{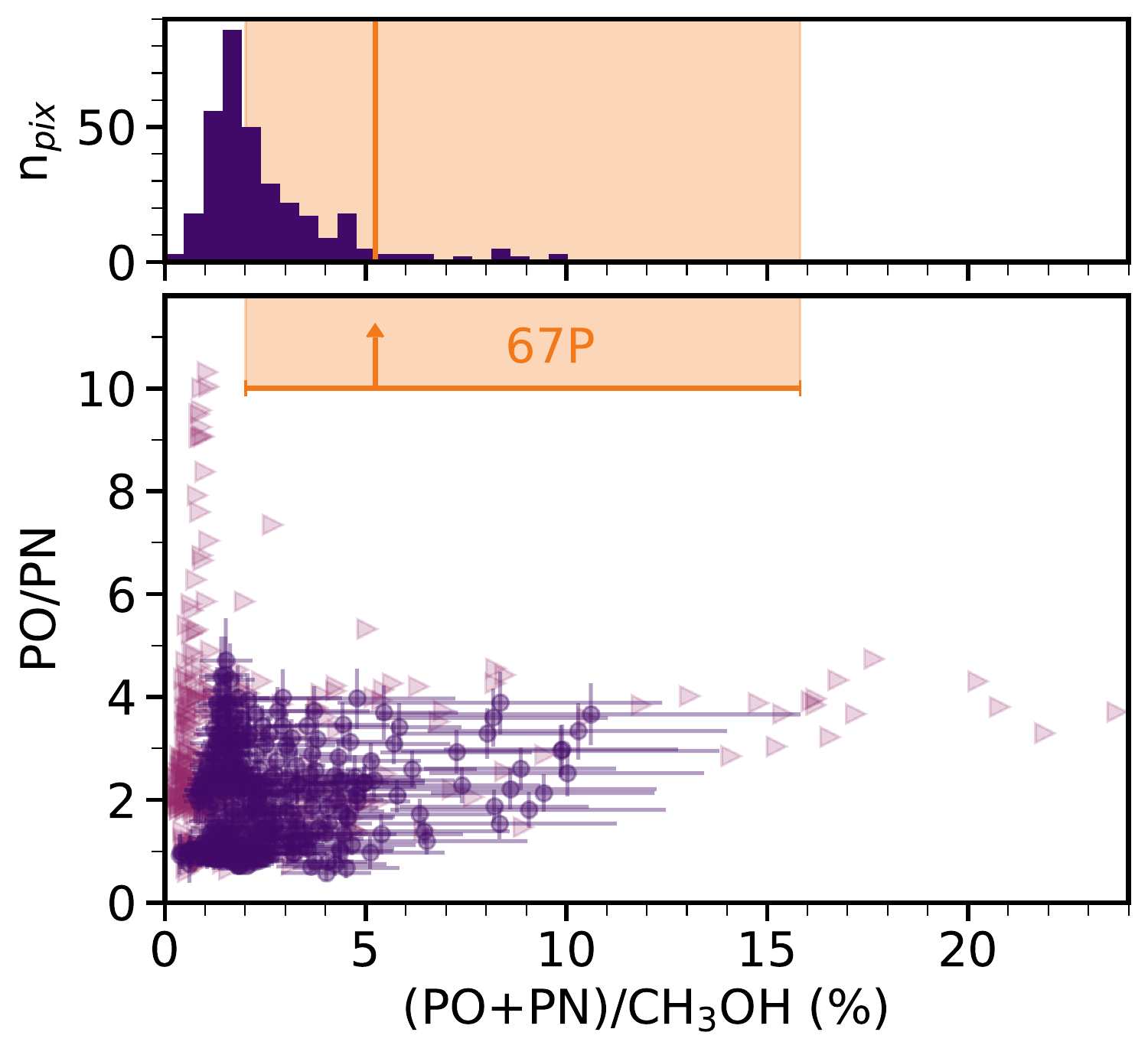}
    \caption{Top: histogram of (PO+PN)/CH$_3$OH ratios in B1-a, including only pixels where $^{13}$CH$_3$OH is detected.  Bottom: (PO+PN)/CH$_3$OH ratios plotted against PO/PN ratios.  Purple markers show pixels in B1-a where $^{13}$CH$_3$OH was detected, and pink triangles show (PO+PN)/CH$_3$OH lower limits.  In both panels, the orange region shows the range of values possible for comet 67P.}
    \label{fig:P_CH3OH_ratio}
\end{figure}

With these limitations in mind, Figure \ref{fig:P_CH3OH_ratio} shows the (PO+PN)/CH$_3$OH column density ratios in B1-a plotted against the PO/PN ratios within individual pixels.  We solve for CH$_3$OH column densities from $^{13}$CH$_3$OH by adopting the local ISM $^{12}$C/$^{13}$C ratio of 68 \citep{Milam2005}.  We include all pixels with well-constrained column densities for both PO and PN.  For pixels in which a $^{13}$CH$_3$OH column density could not be derived due to low SNR, we adopt the 90th percentile column density posterior as the upper limit, and show the (PO+PN)/CH$_3$OH ratio as a lower limit.  Note that this analysis uses PO, PN, and $^{13}$CH$_3$OH images tapered to a circular 2.5" beam.  We recover a similar distribution of PO/PN ratios as was found from the higher-resolution images, though note that several pixels are seen to have even higher PO/PN ratios (up to 10) in the low-resolution data.  The range of possible PO/PN and (PO+PN)/CH$_3$OH ratios in comet 67P are also shown in Figure \ref{fig:P_CH3OH_ratio}, based on the analyses of ROSINA data by \citet{Rubin2019b} and \citet{Rivilla2020}.

In most regions of B1-a where $^{13}$CH$_3$OH is detected, the (PO+PN)/CH$_3$OH ratio is $\sim$1--3\%, with a tail up to 10\%.  In some pixels with CH$_3$OH upper limits, the (PO+PN)/CH$_3$OH lower limit extends up to $\sim$20, though most of (PO+PN)/CH$_3$OH lower limits are not constraining (e.g.~lower limits around 1\%).  By comparison, the bulk (PO+PN)/CH$_3$OH ratio inferred for comet 67P \citep{Rubin2019b} is around 5\%, though a ratio as low as 2\% and as high as 16\% are within the uncertainties.  Just over half of the pixels in B1-a exhibit (PO+PN)/CH$_3$OH ratios consistent within the uncertainties with comet 67P, while the remaining pixels are lower but generally within a factor of $\sim$2.

Figure \ref{fig:P_CH3OH_ratio} also compares the PO/PN ratios measured in B1-a with those inferred for comet 67P.  Only a few pixels in B1-a exhibit PO/PN ratios as high as the lower limit of 10 found in comet 67P.  Indeed, in the higher-resolution analysis no pixels are found with a PO/PN as high as 10 (Figure \ref{fig:PO_PN_ratio}).  $^{13}$CH$_3$OH was not detected at these positions,  so it is unclear at present whether the (PO+PN)/CH$_3$OH ratio is also consistent with that measured in comet 67P.  Even so, for the majority of pixels we find that the PO/PN ratio is lower than that measured towards comet 67P.  We discuss this further in Section \ref{subsec:solarsystem}.

\section{Discussion}
\label{sec:discussion}
\subsection{Origin of phosphorus emission}
\label{subsec:origins}
To date, phosphorus carriers have been detected towards just two low-mass star forming regions: L1157 \citep{Yamaguchi2011, Lefloch2016} and B1-a \citep{Bergner2019}.  While in both cases the emission has been linked to protostellar outflows, the mechanism responsible for releasing phosphorus into the gas remains uncertain.  Our observations (Figure \ref{fig:emission_compare}) reveal a pattern in which the phosphorus molecules emit from the regions where the protostellar outflow (traced by SiO) encounters a dense filament-like structure (traced by CCS).  Thus, we infer that the phosphorus molecules trace high-density ambient material shocked by the outflow.  Indeed, the chemically rich regions observed towards several other outflows, including L1157-B1, seem to require interaction of the outflow with dense, pre-existing clumps \citep{Viti2004}.  

Towards the massive star-forming region AFGL 5142, spatially resolved observations revealed that PO and PN emit from several clumps tracing the the outflow cavity walls \citep{Rivilla2020}.  The clumpy nature of the emission in both B1-a and AFGL 5142 likely arises from a common process, i.e.~an outflow interacting with high-density gas.

The PO and PN line profiles in B1-a are quite narrow, with line widths $<$1 km s$^{-1}$ across the source.  Together with the low excitation temperatures (generally $\sim$10 K), this indicates that the emitting gas is fairly quiescent.  In AFGL 5142, narrow line-widths were interpreted as emission from post-shocked gas \citep{Rivilla2020}, a scenario that is likely true for B1-a as well.  Thus, while shocks appear to play an important role in releasing phosphorus carriers into the gas, the observed phosphorus molecules do not emit directly from the shocked gas.

\citet{Rivilla2020} propose that post-shock photochemistry is a necessary ingredient in the production of PO and PN within the outflow cavity of AFGL 5142.  Given the unusual morphology of the SiO outflow in B1-a, it is not clear whether PO and PN emit from UV-exposed gas such as the ouflow cavity walls.  Observations of a photochemistry tracer such as CN or C$_2$H towards B1-a would help to reveal whether photochemistry is active within the phosphorus emitting regions, or whether their formation does not depend on UV activation.

\subsection{Candidate grain carriers}
\label{subsec:grain_carriers}
PN and PO emit co-spatially with the low-velocity shock tracers $^{13}$CH$_3$OH and SO$_2$, and not co-spatially with the high-velocity shock tracer SiO.  This indicates that prior to the shock event, there is a phosphorus carrier in the solid state that is more volatile than silicate grain. It is unlikely that this parent phosphorus carrier is as volatile as CH$_3$OH, which is commonly detected following ice sublimation in hot cores and hot corinos, whereas the detection of P carriers seems to require shocking \citep{Rivilla2020, Bernal2021}.  Thus, before the shock there seems to be a component of the total phosphorus reservoir that is more refractory than simple ices like CH$_3$OH, and less refractory than silicate grains. 

Of the phosphorus forms that are expected in the Solar Nebula, phosphate minerals (e.g.~apatite, Ca$_5$[PO$_4$]$_3$(F,Cl,OH)) and large phosphorus oxides (e.g.~P$_4$O$_{10}$) should exhibit intermediate volatilities between simple ices like CH$_3$OH and silicate grains \citep[Table \ref{tab:volatility}; ][]{Lodders2003, Pasek2019}.  On the other hand, schreibersite ((Fe,Ni)$_3$P) and metal phases of phosphorus are generally closer in volatility to silicate grains, and thus less compelling candidates to explain the emission patterns we observe in B1-a.  Interestingly, PO bonds are present in both mineral phosphates and phosphorus oxides.  The decomposition of one of these carriers could release PO directly into the gas, thus bypassing the initial stages of gas-phase phosphorus chemistry that are required for models beginning with PH$_3$ \citep[e.g.][]{Jimenez2018}.

Similar to phosphorus, in dense interstellar regions the total abundance of detectable sulfur carriers is generally a small fraction of the Solar sulfur abundance, implying heavy depletion onto grains \citep{Tieftrunk1994}.  The very close correspondence between the emission morphology of SO$_2$ and the phosphorus carriers (especially PN) suggests that a parent sulfur carrier in the grain may have a similar volatility to the parent P carrier.  Candidate refractory S carriers include troilite (FeS) and S polymers (S$_n$).  FeS has a comparable volatility to apatite \citep[Table \ref{tab:volatility}; ][]{Lodders2003}, while the volatility of S$_n$ chains are not well constrained.  Experimental sublimation curves of the small chains S$_2$--S$_4$ show peak desorption temperatures up to $\sim$twice that of H$_2$O ice \citep{Jimenez2011, Mahjoub2017}, and presumably the larger chains continue to decrease in volatility with size.  Thus, both FeS and allotropic sulfur are plausible grain carriers of sulfur with intermediate volatilities between ices and silicates. 

\begin{deluxetable}{lcc}
	\tabletypesize{\small}
	\tablecaption{Approximate volatility sequence of ice/grain species \label{tab:volatility}}
	\tablecolumns{3} 
	\tablewidth{\textwidth} 
	\tablehead{
        \colhead{Molecule}       & 
        \colhead{T$_\mathrm{cond/sub}$ (K)} & 
        \colhead{Ref.}}
\startdata
Water: H$_2$O & 182 & 1\\
Elemental sulfur: S$_2$--S$_4$ & $\sim$170--300 & 2, 3 \\
Phosphorus oxides: P$_4$O$_{9,10}$ & $\lesssim$500 & 4 \\
Troilite: FeS & 704 & 1 \\
Apatite: Ca$_5$[PO$_4$]$_3$F & 739 & 1 \\
Schreibersite: Fe$_3$P & 1248 & 1 \\
Forsterite (silicate): Mg$_2$SiO$_4$ & 1354 & 1 \\
\enddata
\tablenotetext{}{Equilibrium condensation temperatures (Refs. 1,4) or experimental sublimation temperatures (Refs. 2,3) for different ice and grain species considered in this work.  Both T$_\mathrm{cond}$ and T$_\mathrm{sub}$ reflect a molecule's volatility, however we note that this comparison is approximate because of the different methods and pressures used in different works.  References: [1] \citet{Lodders2003}, [2] \citet{Jimenez2011}, [3] \citet{Mahjoub2017}, [4] \citet{Pasek2019}.}
\end{deluxetable}

\subsection{Relationship between PO and PN}
The PO/PN ratios measured in B1-a vary from $\sim$1--8.  This agrees well with the range derived from spatially-resolved observations of AFGL 5142 \citep{Rivilla2020}.  The median ratio across the source is 1.9, which is comparable to the source-averaged PO/PN ratios of $\sim$1--3 measured previously towards B1-a \citep{Bergner2019} as well as towards other dense ISM sources \citep{Lefloch2016, Rivilla2016, Rivilla2018, Bernal2021}.  Still, the detection of high PO/PN ratios in some regions of B1-a, as well as potential correlations between the PO/PN ratio and the local physical conditions (Figure \ref{fig:lw_columns}), illustrate the need for spatially resolved observations of phosphorus molecules in other sources to better constrain the factors driving PO and PN formation.
 
The relative intensities of PN and PO within the N and S clumps are notably different: PO emits only very weakly from the N clump, while PN is comparably bright towards both the N and S clumps (Figure \ref{fig:obs_summary}).  Correspondingly, the PO/PN ratio is relatively low in most of the N clump, and higher in the S clump.  Based on phosphorus chemical models described in the literature, the relatively weak emission of PO in the N clump could be due to (i) preferential photodissociation of PO in a moderate UV field, (ii) a low gas-phase O abundance, or (iii) a time-dependent consumption of PO.  

\begin{figure}
\centering
    \includegraphics[width=\linewidth]{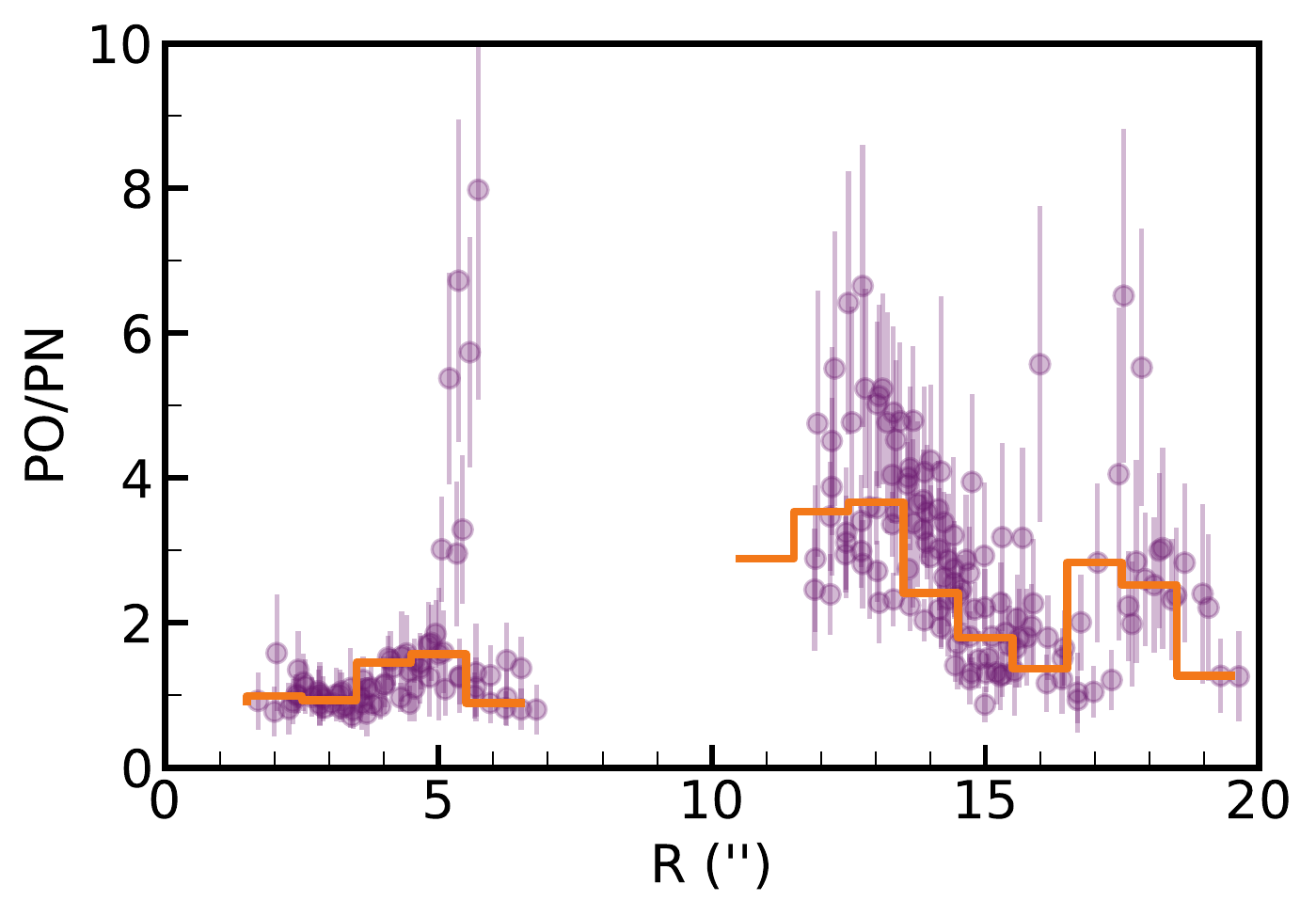}
    \caption{PO/PN as a function of the projected distance from the protostar.  Individual pixels are shown as purple markers, and orange lines represent the median within 1$\arcsec$ bins.}
    \label{fig:popn_vs_r}
\end{figure}

(i) PO is predicted to have a higher photodissociation rate than PN \citep{Jimenez2018}.  Selective PO dissociation should become important when the UV field is strong enough for photodissociation to be active, but not so strong that all molecules are photodissociated.  This mechanism was proposed to explain the increasing PO/PN ratio with distance from the millimeter core observed in the massive star-forming region AFGL 5142 \citep{Rivilla2020}.  While we indeed see a higher median PO/PN ratio in the S clump compared to the N clump, there is not a straightforward trend between the PO/PN ratio and radius from the protostar in B1-a (Figure \ref{fig:popn_vs_r}), as was seen in AFGL 5142.  While projection effects could be masking an underlying trend with distance, at present it remains unclear if photochemistry is an important player in the phosphorus chemistry within B1-a.

(ii) The different PO/PN ratios in the N vs.~S clumps could arise from a difference in the gas-phase O abundance (or, O/N ratio).  Models predict that PO forms efficiently in the gas-phase through PH + O $\rightarrow$ PO + H, and possibly P + OH $\rightarrow$ PO + H.  \citep{Aota2012, Lefloch2016, Jimenez2018, Garcia2021}.  PN can subsequently form via PO + N $\rightarrow$ PN + O.  This chemical sequence was proposed by both \citet{Rivilla2020} and \citet{Bernal2021} to explain the formation of PO and PN towards AFGL 5142 and Orion-KL.  In this scenario, the gas-phase O and N abundances should regulate how efficiently PO and PN are formed from PH or P.  Previous explanations for a low O/N ratio (and low PO/PN ratio) include preferential freeze-out of O-bearing species in cold regions \citep{Rivilla2020}, or a high efficiency of shock-induced sputtering of NH$_3$ into the gas \citep{Lefloch2016}.  Alternatively, we propose that a higher gas-phase O abundance could be related to a higher efficiency of ice desorption.  $^{13}$CH$_3$OH emission is, like PO, much weaker in the N clump compared to the S clump (Figure \ref{fig:emission_compare}).  If H$_2$O desorbs similarly to CH$_3$OH ice, then we expect the gas-phase O abundance to be significantly higher in the S clump than the N clump.  Causes of differential ice desorption between the N and S clumps could include different shock velocities at the two positions, or a time-dependent re-adsorption of ice species post-shock.  Observations of additional N and O carriers towards B1-a are needed to elucidate if the N clump, characterized by a lower PO/PN ratio, shows systematically lower O/N ratios.

(iii) If PO is directly released into the gas following decomposition of a semi-refractory phosphate mineral or phosphorus oxide (Section \ref{subsec:grain_carriers}), it is possible that the PO/PN ratio reflects the time since PO was released into the gas.  As noted above, PN can form in the gas via reactions of PO + N.  Meanwhile, models struggle to reproduce PO/PN ratios$>$1 with the canonical phosphorus chemical network \citep{Jimenez2018}.  Thus, it may be that PO released into the gas is converted to PN, but not efficiently re-formed, resulting in a decrease in PO/PN with time post-shock.

It is also interesting to compare the emission morphologies of PO and PN with other low-velocity shock tracers (Figure \ref{fig:emission_compare}).  As noted in Section \ref{subsec:outflow_tracers}, CH$_3$OH is expected to be a direct product of ice sputtering, while SO$_2$ may result from gas-phase chemistry following sputtering of a different solid S carrier \citep{Bachiller1997,Jimenez2005}.  The similarity in morphology of PO with CH$_3$OH and PN with SO$_2$ suggests that PO is more directly related to the grain P carrier compared to PN.  This supports a scenario in which PN is formed after PO, i.e.~via the gas-phase reaction PO+N.

The higher PO/PN ratios observed in regions with narrower line widths (Figure \ref{fig:lw_columns}) may, in principle, encode information about the conditions favoring PO vs.~PN formation.  However, given the present constraints it is unclear whether this trend is causal (e.g.~PO formation is favored in more quiescent regions) or coincidental.  Indeed, the narrower line-widths in the S clump could be due to numerous factors, for instance if the shock at the S clump was slower in the first place, or if it is older and has dissipated more.  Better constraints on the outflow shock physics in the N vs.~S clumps, e.g.~through multi-line CO observations, are needed to interpret this trend. 

\subsection{Comparison to comet 67P}
\label{subsec:solarsystem}
As a low-mass protostar embedded in an active region of star formation, B1-a represents an analog to the proto-Sun.  Thus, a comparison with the Solar System comet 67P offers insight into how volatile phosphorus was incorporated into the comet.  We emphasize that measurements of volatile phosphorus molecules exist towards only one comet and two low-mass protostars, and expanded demographics are greatly needed to improve our understanding of phosphorus inheritance and evolution.

As discussed in Section \ref{subsec:grain_carriers}, prior to protostellar outflow shocking, the parent phosphorus carrier in the solid state seems to have an intermediate volatility between simple ices and silicate grains.  The PO detected in comet 67P is therefore unlikely to be an ice constituent from before the protostellar stage, given its high volatility \citep{Wakelam2012}.  Outflow shocks during the protostellar stage appear to be an important step in producing volatile phosphorus carriers like PO and PN, which can condense into the ice phase and may subsequently be incorporated into cometary building blocks.  Indeed, the (PO+PN)/CH$_3$OH ratios in B1-a (generally $\sim$1--3\%) are similar to the ratio measured in comet 67P (5\%).  This implies a comparable reservoir of volatile phosphorus in both objects, consistent with a scenario in which the volatile P in cometary ices is sourced from protostellar shocks.

The PO/PN ratio is generally lower in B1-a by a factor of a few compared to the lower limit measured for comet 67P.  Figure \ref{fig:lw_columns} shows that the PO/PN ratio in B1-a is highest in the regions with the narrowest line-widths.  If freeze-out is more efficient in this cool/quiescent gas, the material that re-freezes onto grains could be preferentially enriched in PO.  Thus, it may be that the ice-phase PO/PN ratio incorporated into comets is higher than that observed in most of the protostellar gas.  Alternatively, additional phosphorus processing may happen in the ice phase that shifts the PO/PN ratio between the protostellar and comet-forming phases.  

\section{Conclusions}
\label{sec:concl}

Our ALMA observations of PO and PN towards B1-a are the first spatially resolved images of phosphorus carriers towards a protosolar analog.  Our main conclusions are as follows.

\begin{enumerate}[topsep=2pt,itemsep=2pt,partopsep=1ex,parsep=1ex, leftmargin=20pt]
    \item PO and PN do not emit from the outflow as traced by SiO, but from two distinct clumps where the outflow encounters a dense filament traced by CCS.  The low-velocity shock tracers $^{13}$CH$_3$OH and SO$_2$ also emit from these clumps.  Thus, weak shocking of dense ambient gas by the outflow appears to be responsible for releasing P into the gas.
    
    \item The observed gas-phase PO and PN are likely daughter products of a solid phosphorus reservoir with an intermediate volatility between ices and silicate grains.  Phosphate minerals or phosphorus oxides are good candidates for a moderately refractory P reservoir.  In this scenario, interstellar shocks may play an important role in converting semi-refractory phosphorus to more volatile forms like PO and PN, which may be incorporated into cometary ices \citep[e.g.][]{Altwegg2016}.
    
    \item The PO/PN ratio varies from $\sim$1--8 across B1-a, and is generally lower in the northern clump than the southern clump, signifying a distinct shock chemistry/physics in the two clumps.  Causes of the weak PO emission in the northern clump may include selective PO photodissociation, a lower gas-phase O/N ratio, or a time-dependent conversion of PO to PN.
    
    \item The volatile phosphorus abundance traced by (PO+PN)/CH$_3$OH is similar in B1-a and comet 67P.  However, B1-a generally exibits lower PO/PN ratios than the comet, which may reflect preferential freeze-out of PO-rich gas, ice-phase phosphorus chemistry between the protostellar and disk stages, or an intrinsically different O/N ratio between B1-a and the proto-solar nebula.
    
\end{enumerate}  

The spatial information encoded in these observations has provided a significantly more detailed view of the phosphorus chemistry compared to previous single-dish observations.  Still, our understanding of interstellar phosphorus chemistry is currently limited by (i) the small number of volatile P carriers detected towards star-forming regions, (ii) a poor understanding of the B1-a outflow physics, and (iii) an extremely limited sample size of low-mass protostars with detections of phosphorus carriers.  Future observational efforts addressing these limitations are required to more fully constrain the astrochemical inheritance and evolution of this key prebiotic element.

\acknowledgments 

This paper makes use of the following ALMA data: ADS/JAO.ALMA\#2019.1.00708.S. ALMA is a partnership of ESO (representing its member states), NSF (USA) and NINS (Japan), together with NRC (Canada), MOST and ASIAA (Taiwan), and KASI (Republic of Korea), in cooperation with the Republic of Chile. The Joint ALMA Observatory is operated by ESO, AUI/NRAO and NAOJ.  J.B.B. acknowledges support from NASA through the NASA Hubble Fellowship grant \#HST-HF2-51429.001-A awarded by the Space Telescope Science Institute, which is operated by the Association of Universities for Research in Astronomy, Incorporated, under NASA contract NAS5-26555. 
K.I.\"O. acknowledges support from the Simons Foundation (SCOL \#321183).

\software{
{\fontfamily{qcr}\selectfont CASA} \citep{McMullin2007},
{\fontfamily{qcr}\selectfont Matplotlib} \citep{Hunter2007},
{\fontfamily{qcr}\selectfont NumPy} \citep{VanDerWalt2011},
{\fontfamily{qcr}\selectfont Radex} \citep{vanderTak2007},
{\fontfamily{qcr}\selectfont Astropy} \citep{Astropy2013},
{\fontfamily{qcr}\selectfont Scipy} \citep{SciPy2020},
{\fontfamily{qcr}\selectfont kalepy} \citep{Kelley2021}.
}

\clearpage

\appendix 
\FloatBarrier
\section{Observational details}
\label{sec:app_obsdeets}
Table \ref{tab:obs_log} summarizes the Band 3 and Band 4 observations taken with ALMA.  Correlator configurations for the spectral windows containing the line targets used in this work are shown in Table \ref{tab:spec_setups}.  The image parameters for PO and PN, along with integrated flux densities and emitting areas, are listed in Table \ref{tab:image_dat}.  

\begin{deluxetable*}{lcccccrcc}
	\tabletypesize{\footnotesize}
	\tablecaption{Observation details \label{tab:obs_log}}
	\tablecolumns{9} 
	\tablewidth{\textwidth} 
	\tablehead{
	    \colhead{ID}       & 
		\colhead{Config.}       & 
		\colhead{\# Ant.} &
        \colhead{Baselines}       & 
		\colhead{Time On-source} & 
		\colhead{\# EB} &
		\colhead{Dates}       &
		\colhead{Bandpass \& Flux Cal.} &
		\colhead{Phase Cal.} \\
	    \colhead{}       & 
		\colhead{}       & 
		\colhead{} &
        \colhead{(m)}       & 
		\colhead{(min.)} & 
		\colhead{} &
		\colhead{}       &
		\colhead{} &
		\colhead{} 
		}
\startdata
B3 hi-res & 12m C43-4 & 54 & 15--1231 & 144 & 4 & Mar. 08--18 2020 & J0510+1800 & J0366+3218 \\
B3 lo-res & 12m C43-1 & 45 & 15--313  &  37 & 1 & Dec. 19 2019 & J0423-0120 & J0366+3218 \\
B4 hi-res & 12m C43-3 & 43 & 15--783  &  70 & 2 & Mar. 02 2020 & J0237+2848, J0510+1800 & J0366+3218 \\
B4 lo-res & 7m        & 11 &  9--49   & 111 & 3 & Nov. 22--25 2019 & J0423-0120 & J0366+3218 
\enddata
\tablenotetext{}{}
\end{deluxetable*}

\begin{deluxetable*}{lccc}
	\tabletypesize{\footnotesize}
	\tablecaption{Spectral setups \label{tab:spec_setups}}
	\tablecolumns{4} 
	\tablewidth{\textwidth} 
	\tablehead{
		\colhead{Center frequency (GHz)}       & 
		\colhead{Target molecules } &
        \colhead{Bandwidth (MHz)}       & 
		\colhead{Spectral resolution (kHz / km s$^{-1}$)} 
		}
\startdata
\multicolumn{4}{c}{Band 3} \\
\hline 
93.979745 & PN, CCS & 234.38 & 141 / 0.45 \\
108.734849 & PO & 468.75 & 282 / 0.78 \\
108.998445 & PO & 58.59 & 141 / 0.39 \\
109.060000 & PO & 58.59 & 141 / 0.39 \\
109.190000 & PO & 58.59 & 141 / 0.39 \\
109.271189 & PO & 58.59 & 141 / 0.39 \\
\hline
\multicolumn{4}{c}{Band 4} \\
\hline 
140.306170 & SO$_2$ & 58.59 & 141  / 0.30 \\
140.967661 & PN & 117.19 & 141 / 0.30  \\
141.615000 & $^{13}$CH$_3$OH & 117.19 & 141 / 0.30 \\
152.700000 & PO & 117.19 & 141 / 0.28 \\
152.855454 & PO & 117.19 & 141 / 0.28 \\
\enddata
\tablenotetext{}{}
\end{deluxetable*}

\begin{deluxetable*}{lcclrc}
	\tabletypesize{\small}
	\tablecaption{PO \& PN image parameters \label{tab:image_dat}}
	\tablecolumns{6} 
	\tablewidth{\textwidth} 
	\tablehead{
        \colhead{Spectral window}       & 
        \colhead{Beam Dim.} &
		\colhead{Chan. rms$^a$} & 
		\colhead{Lines} &
		\colhead{Int. flux density$^b$} & 
	    \colhead{$\Omega^c$} \\
	    \colhead{} & 
	    \colhead{ ($\arcsec \times \arcsec$) }& 
	    \colhead{(mJy beam$^{-1}$)} & 
	    \colhead{(GHz)} & 
	    \colhead{(Jy beam$^{-1}$ km s$^{-1}$)} &
	    \colhead{(square $\arcsec$)} }
\startdata
PN B3   & 1.5 $\times$ 0.85 & 1.6 & 93.980 & 11.78 [0.27] & 28.9\\ 
PN B4   & 1.5 $\times$ 0.85 & 2.5 & 140.968 & 30.40 [0.39] & 40.2\\ 
\hline
PO B3-1 & 1.5 $\times$ 0.85 & 1.8 & 108.998 & 5.03 [0.10] & 18.7\\ 
PO B3-2 & 1.5 $\times$ 0.85 & 1.8 & 109.045 & 2.68 [0.06] & 12.8\\ 
PO B3-3 & 1.5 $\times$ 0.85 & 1.8 & 109.206 & 5.72 [0.11] & 20.5\\ 
PO B3-4 & 1.5 $\times$ 0.85 & 1.8 & 109.271 & 0.22 [0.02] & 1.2\\ 
        & & & 109.281 & 2.68 [0.07] & 13.2\\ 
PO B4-1 & 1.5 $\times$ 0.85 & 2.8 & 152.657 & 11.67 [0.19] & 28.3\\ 
        & & & 152.680 & 6.41 [0.18] & 18.2\\ 
PO B4-2 & 1.5 $\times$ 0.85 & 2.8 & 152.855 & 12.53 [0.20] & 29.0\\ 
        & & & 152.888 & 8.02 [0.18] & 23.0\\ 
\enddata
\tablenotetext{}{$^a$In velocity bins of 0.4 km s$^{-1}$ for Band 3 and 0.3 km s$^{-1}$ for Band 4.  $^b$Integrated over 4.5--6.6 km s$^{-1}$ for PO and 2.9--7.3 km s$^{-1}$ for PN, including only pixels with a SNR$>$3 in the moment zero map.  $^c$Solid angle of the emitting region.}
\end{deluxetable*}

\FloatBarrier

\begin{figure*}
\centering
    \includegraphics[width=0.9\linewidth]{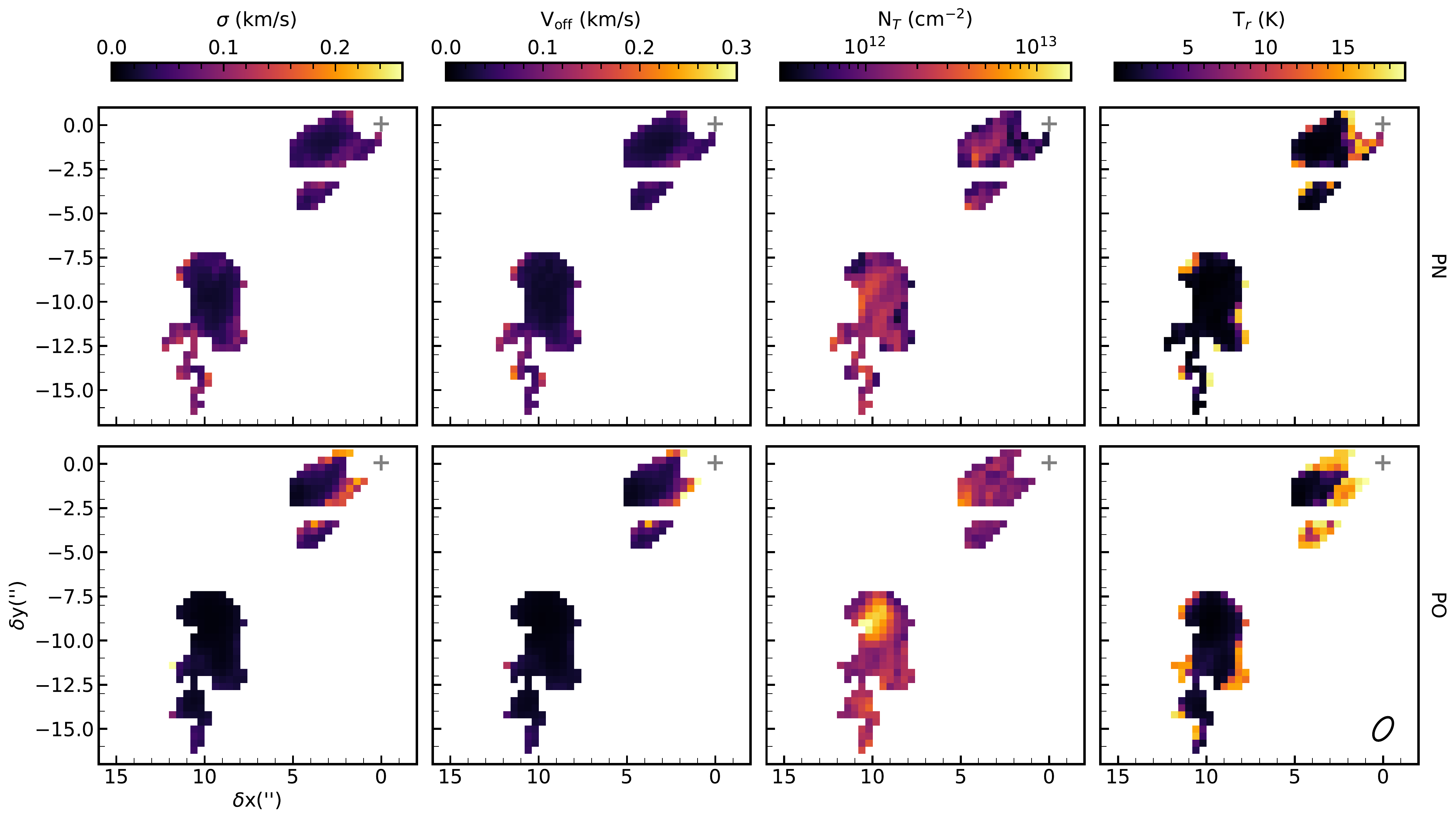}
    \caption{Maps of the mean 1$\sigma$ parameter uncertainties (i.e. 50-16th and 84-50th percentile posteriors) from spectral line fitting of PN (top) and PO (bottom).  Only pixels with $>$3$\sigma$ constraints on the column density are shown.  The restoring beam (common to PN and PO) is shown in the bottom right panel.}
    \label{fig:fit_param_errors}
\end{figure*}
\section{Spectral line fitting}
\label{sec:app_linefits}
With multi-line coverage of PN, PO, and $^{13}$CH$_3$OH, we performed spectral line fitting to constrain the molecular column densities and rotational temperatures.  Section \ref{sec:app_lte} describes our local thermodynamic equilibrium (LTE) fitting routine to produce column density maps.  We also tested a non-LTE fitting routine on several representative spectra to explore the possible role of non-LTE effects, described in Section \ref{sec:app_nonlte}.

\subsection{LTE routine}
\label{sec:app_lte}
For the LTE fitting of PN, PO, and $^{13}$CH$_3$OH, we fit all observed spectra simultaneously for the four free parameters of total column density ($N_T$), the rotational temperature ($T_r$), the Gaussian line width $\sigma$, and a velocity offset $V_\mathrm{off}$.  

We produce synthetic spectra by first solving for the optical depth at each line center, $\tau_0$, using:
\begin{equation}
    \tau_{0} = \frac{N_T}{Q(T_{r})} e^{-E_{u}/T_{r}} \frac{g_{u} A_{u}c^3}{8\pi\nu^3}\frac{1}{\sigma \sqrt{2\pi}} (e^{h\nu/kT_{r}}-1),
    \label{eq:tau}
\end{equation}
where $Q(T_{r})$ is the molecular partition function, $E_{u}$ is the upper state energy in K, $g_{u}$ is the upper state degeneracy, $A_{u}$ is the Einstein coefficient and $\nu$ is the line frequency.  The optical depth profile as a function of velocity $V$ is then found from:
\begin{equation}
    \tau_\nu = \tau_{0} \mathrm{exp}\Big{(}\frac{-(V- V_\mathrm{off})^2}{2\sigma^2}\Big{)},
    \label{eq:tau_nu}
\end{equation}
We then produce a synthetic intensity profile using:
\begin{equation}
    I_\nu = [B_\nu(T_{r}) - B_\nu(T_{bg})] \times (1-e^{-\tau_v}) \times \Omega,
    \label{eq:F_nu}
\end{equation}
where $B_\nu$ is the Planck function and $\Omega$ is the restoring beam solid angle.  The continuum brightness co-spatial with the phosphorus molecule emission is small, and we adopt the cosmic microwave background of 2.73 K as the background temperature $T_{bg}$.  The spectral line parameters used for fitting are listed in Table \ref{tab:line_dat}.

In fitting the data, we include a 10\% flux uncertainty to account for the ALMA calibration uncertainty.  We sample the posterior distributions of the fit parameters $N_T$, $T_r$, $\sigma$, and $V_\mathrm{off}$ using the affine-invariant Markov Chain Monte Carlo package \texttt{emcee} \citep{Foreman2013}.  

To derive column density maps, we fitted only pixels for which the PN or PO Band 4 velocity-integrated intensity map has an SNR$>$5.  We consider a molecule undetected within a given pixel if the median value of the column density posterior distribution is $<3\times\sigma_l$, where $\sigma_l$ is the lower uncertainty as found from the 16th percentile posterior.  In this case, we instead report a column density upper limit based on the 90th percentile posterior.  The resulting maps of the best-fit parameters are shown in Figure \ref{fig:fit_parameters}, and the parameter uncertainty maps are shown in Figure \ref{fig:fit_param_errors}.

\subsection{Non-LTE routine}
\label{sec:app_nonlte}

We also tested a non-LTE fitting routine in order to explore the robustness of our LTE fits.  For this, we used a non-LTE spectral model based on RADEX \citep{vanderTak2007}, along with the PO and PN collisional rates provided by \citet{Lique2018} and \citet{Tobola2007}.  We fit for the column density ($N_T$), kinetic temperature ($T_\mathrm{kin}$), line width ($\sigma$), velocity offset ($V_\mathrm{off}$), and gas density (n$_H$).  We fitted spectra extracted from a 5$\times$5 pixel box centered on the N and S positions marked in Figure \ref{fig:obs_summary}.  We also fitted the same spectra with our LTE routine to compare the resulting column densities.

\begin{figure*}
\centering
    \includegraphics[width=\linewidth]{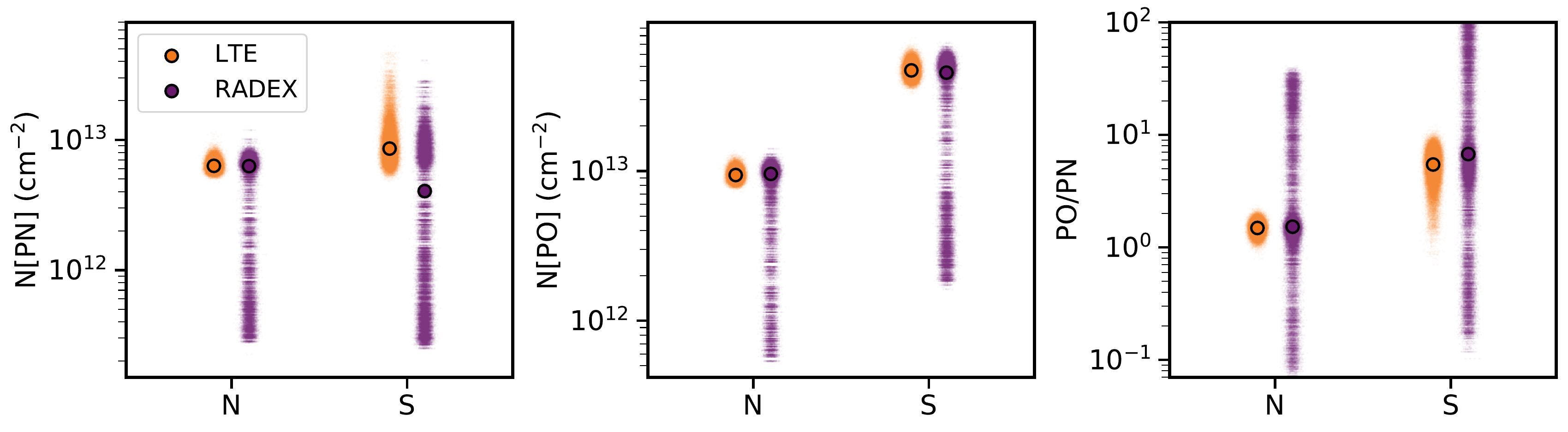}
    \caption{Carpet plots \citep{Kelley2021} show the PN, PO, and PO/PN column density posteriors derived from LTE and RADEX analyses, for spectra extracted from the N and S emission peaks.  Black circles represent the median posterior values.}
    \label{fig:lte_nonlte_compare}
\end{figure*}

\begin{figure*}
\centering 
\includegraphics[width=.5\linewidth]{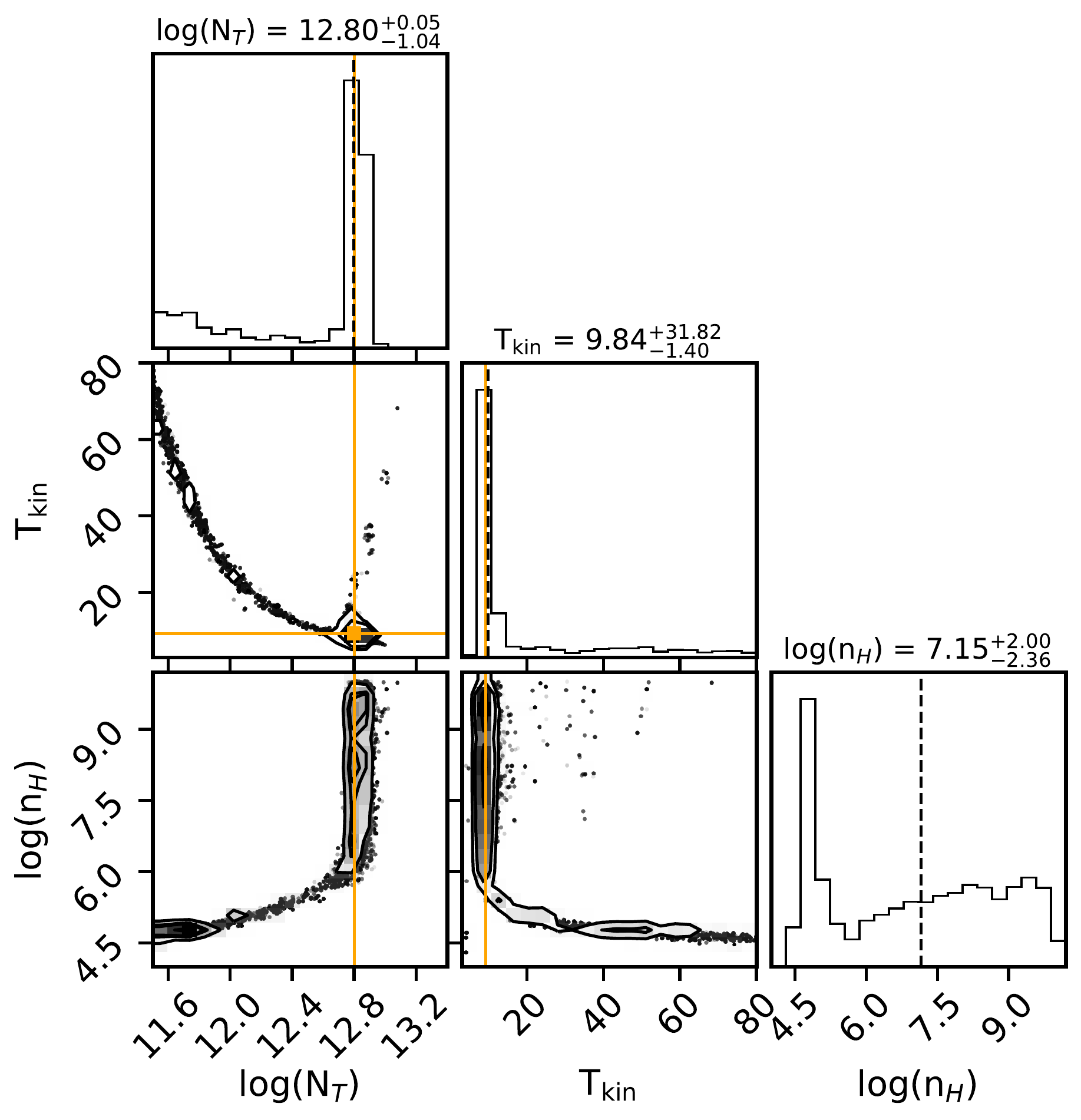}\includegraphics[width=.5\linewidth]{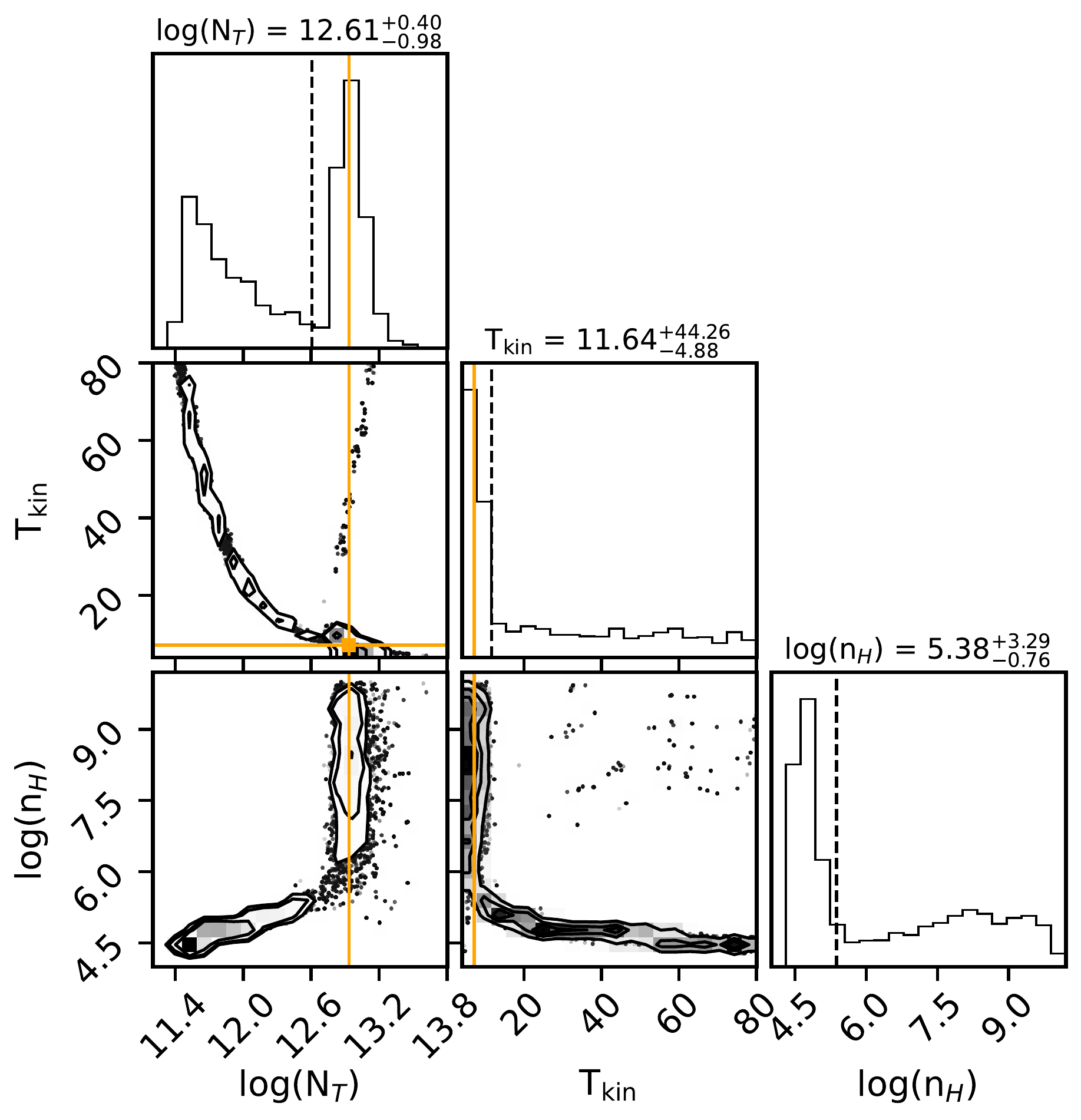}
         \caption{Corner plots \citep{corner2016} from RADEX fitting of PN spectra extracted from the N (left) and S (right) emission peaks.  For clarity, only the column density (cm$^{-2}$), kinetic temperature (K), and gas density (cm$^{-3}$) fits are included.  Black dashed lines show the median posterior values, and for comparison the solid orange lines show the LTE-derived median posteriors.}
         \label{fig:pn_corners}
\end{figure*}

Figure \ref{fig:lte_nonlte_compare} shows the PN, PO, and PO/PN column density posteriors for the RADEX vs.~LTE treatment.  The median posteriors are generally in good agreement between the LTE and RADEX fits.  The largest difference is seen for PN at the S position, which is a factor of $\sim$2 lower for the RADEX fit than for the LTE fit.  While we see good agreement in the median posterior values, the RADEX fits show a much larger range in column densities posteriors than the LTE fit.  This reflects that there are essentially two allowable solutions for the RADEX line fits: a higher column density corresponding to a higher emitting gas density (i.e.~LTE), or a lower column density corresponding to a lower emitting gas density (i.e.~non-LTE).  This can be seen clearly in Figure \ref{fig:pn_corners}, which shows the corner plots for the RADEX PN fits. At both the N and S position, there is some degeneracy between the high and low column density solutions.  A similar behavior is seen for the PO fits.

Thus, this analysis demonstrates that with only two J-level transitions, there are insufficient data to unambiguously constrain the emitting conditions and column densities of PN and PO. At the same time, the LTE solution is favored over the non-LTE solution for the cases tested here, reflected by the good agreement between the median column density posteriors of the RADEX vs.~LTE fits.  A similar outcome was seen in the MCMC non-LTE fitting of PN 3--2 and 2--1 by \citet{Haasler2021}.  We therefore use the LTE fitting routine for the main analysis presented in this paper, with the caveat that non-LTE solutions remain a possibility.  Additional coverage of higher-J PN and PO lines is needed to firmly constrain the PN and PO emitting conditions and column densities.

\FloatBarrier
\clearpage
\bibliography{references}

\end{document}